\documentclass[12pt,preprint]{emulateapj}
\usepackage{graphicx}
\usepackage{epsfig}
\usepackage{lscape}

\def\degree{\ensuremath{^\circ}}
\def\spi{{\it Spitzer}}
\begin{document}

\title{Mid-infrared spectra of optically selected type 2 quasars}

\author{
Nadia L. Zakamska$^{1,2}$,
Laura G\'omez$^{3,4}$,
Michael A. Strauss$^5$,
Julian H. Krolik$^6$}
\affil{1 - Institute for Advanced Study, Einstein Dr., Princeton, NJ 08540; 2 - \spi\ Fellow; John N. Bahcall Fellow; \\
3 - Max-Planck-Institut f\"ur Radioastronomie, Auf dem H\"ugel 69, D-53121 Bonn Germany; \\
4 - Centro de Radioastronom\'{\i}a y Astrof\'{\i}sica, UNAM, Apartado Postal 3-72, Morelia, Michoac\'an 58089, Mexico; \\
5 - Princeton University Observatory, Princeton, NJ 08544; \\
6 - Department of Physics and Astronomy, Johns Hopkins University, 3400 North Charles Street, Baltimore, MD 21218-2686}

\begin{abstract}
Type 2 quasars are luminous Active Galactic Nuclei whose central engines are seen through large amounts of gas and dust. We present \spi\ spectra of twelve type 2 quasars selected on the basis of their optical emission line properties. Within this sample, we find a surprising diversity of spectra, from those that are featureless to those showing strong PAH emission, deep silicate absorption at 10\micron, hydrocarbon absorption, high-ionization emission lines and H$_2$ rotational emission lines. About half of the objects in the sample are likely Compton-thick, including the two with the deepest Si absorption. The median star-formation luminosity of the objects in our sample measured from the strength of the PAH features is $5\times 10^{11}L_{\odot}$, much higher than for field galaxies or for any other AGN sample, but similar to other samples of type 2 quasars. This suggests an evolutionary link between obscured quasars and peak star formation activity in the host galaxy. Despite the high level of star formation, the bolometric output is dominated by the quasar in all cases. For a given strength of 10\micron\ Si absorption, ULIRGs are significantly colder than are type 2 quasars (their $F_{\nu}[14.5\micron]/F_{\nu}[27.5\micron]$ ratio is 0.5 dex lower), perhaps reflecting different obscuration geometries in these sources. We find that the appearance of the 10\micron\ feature (i.e., whether it shows in emission or in absorption) is well-correlated with the optical classification in type 1 and type 2 quasars, contrary to some models of clumpy obscuration. Furthermore, this correlation is significantly stronger in quasars ($L_{\rm bol}\ga 10^{45}$ erg/s) than it is in Seyfert galaxies ($L_{\rm bol}\ll 10^{45}$ erg/s).
\end{abstract}


\section{Introduction}
Type 2 quasars are powerful ($L_{\rm bol}\ga 10^{45}$ erg/s) active galactic nuclei (AGNs) whose central regions are shielded from the observer by large amounts of gas and dust. Until recently, the existence of dust-enshrouded quasars was in question, but now these objects are routinely discovered at X-ray (e.g., \citealt{szok04, netz06}), infrared \citep{lacy04, ster05, mart06} and optical \citep{zaka03, reye08} wavelengths. A significant controversy remains regarding their space density and contribution to the AGN population luminosity budget. Some analyses suggest that there are at least as many obscured as unobscured quasars even at the highest luminosity end \citep{reye08, mart06, poll08}, whereas others argue for a strong decline of obscured quasar fraction with luminosity \citep{ueda03, maio07, trei08}. 

If an AGN is entirely embedded in an optically thick dusty cloud, virtually all of its original radiation (emitted primarily at optical to soft X-ray wavelengths) is absorbed by dust and is then re-emitted thermally in the IR. In this case, there is no optical signature of the AGN, but its presence may be revealed when hard X-rays and a large IR luminosity are detected. In this way, some ultra-luminous infrared galaxies (ULIRGs) have been shown to contain embedded AGNs \citep{risa00, fran03, teng05}. The dominant energy source in ULIRGs remains a matter of controversy. Some authors argue that they are powered by star formation, with the AGN contributing only a small fraction \citep{fran03}, while others suggest that buried AGNs are energetically dominant in many ULIRGs \citep{iman07}. 

An alternative possibility is that the nucleus is only partially covered by the obscuring material. In this case, the AGN is seen as obscured or unobscured depending on the line of sight, making its UV, optical and X-ray emission highly anisotropic \citep{anto93}. In particular, UV and X-ray emission can make its way out through the holes in the obscuring cloud and photo-ionize extended regions, which in turn produce powerful optical emission lines. Using the Sloan Digital Sky Survey (SDSS; \citealt{york00, stou02, adel08}) we have selected hundreds of type 2 quasars at redshifts $z<0.8$ on the basis of their optical emission line properties \citep{zaka03, reye08}.

IR observations are key to determining the physical conditions of the obscuring material and its geometric properties such as its covering factor and extent. In this paper we present \spi\ Space Telescope \citep{wern04} Infrared Spectrograph (IRS; \citealt{houc04}) observations of twelve type 2 quasars from the SDSS sample. In Section \ref{sec_data} we describe our sample selection and data reduction. In Section \ref{sec_spectra} we present a detailed analysis of the spectral features. We compare our sample with other samples of type 2 quasars in Section \ref{sec_compare}, we discuss our results in the context of radiative transfer models in Section \ref{sec_models} and we conclude in Section \ref{sec_conc}. 

Throughout the paper, we adopt a cosmology with $h=0.7$, $\Omega_m=0.3$, $\Omega_{\Lambda}=0.7$. Objects are identified as SDSS Jhhmmss.ss$+$ddmmss.s in Figure \ref{pic1} and Table \ref{tab_summary} and are shortened to SDSS Jhhmm$+$ddmm in the text. We refer to type 1 and type 2 AGNs based on their optical classification (in type 1 objects, broad lines are detected and in type 2, they are not). When discussing the luminosity dependence of AGN phenomenon, we distinguish between Seyfert galaxies ($L_{\rm bol}\ll10^{45}$ erg/s) and quasars ($L_{\rm bol}\ga 10^{45}$ erg/s) based on their bolometric luminosities. 

\section{Sample selection, observations and data reduction}
\label{sec_data}

The type 2 quasars discussed in this paper were selected from the spectroscopic database of the SDSS on the basis of their optical emission line properties \citep{zaka03, hao05a, reye08}. Briefly, type 2 AGNs are required to show narrow emission lines with ratios characteristic of a non-stellar ionizing continuum. We then use the luminosity of the [OIII]5007\AA\ emission line as a proxy for the intrinsic luminosity of the obscured nucleus \citep{zaka03} and classify type 2 AGNs as quasars if  $L$[OIII]$>10^{8.5}L_{\odot}$ and as Seyfert galaxies otherwise. \citet{reye08} obtained a sample of 887 type 2 quasars at redshifts $0<z<0.8$ by applying these criteria to the most recent SDSS data. [OIII] luminosities were not corrected for reddening, and they are thus likely to be underestimated \citep{reye08}.

Our extensive follow-up work has demonstrated that the objects that we classified as type 2 quasars based on their optical properties indeed fit operational definitions of type 2 quasars used at other wavelengths. In particular, they typically have high X-ray luminosities, $L_{2-10 {\rm keV}}>10^{44}$ erg/s, and large amounts of neutral gas along the line of sight, $N_H>10^{22}$ cm$^{-2}$ \citep{zaka04, ptak06, vign04, vign06}. Furthermore, using optical spectropolarimetry, we demonstrated that at least some of the objects in our sample harbor broad-line (type 1) AGNs in their centers \citep{zaka05}.

In Cycle 1 of the \spi\ Space Telescope we conducted seven-band IRAC+MIPS photometry of 25 type 2 quasars from \citet{zaka03} and \citet{hao05a}. These objects were selected primarily based on their [OIII]5007\AA\ line luminosity (specifically, $L$[OIII]$>10^9L_{\odot}$). \spi\ photometric observations (Zakamska et al., in prep.), as well as archival IRAS data \citep{zaka04}, show that type 2 quasars from our sample typically have IR luminosities in excess of $10^{45}$ erg/s, placing them into the classical quasar luminosity regime and validating our [OIII]5007\AA-based luminosity criterion. We selected ten sources from this sample with the highest IRAC-8\micron\ and MIPS-24\micron\ fluxes ($F_{\nu}[8\micron]>1.5$ mJy and $F_{\nu}[24\micron]>6$ mJy) for spectroscopic follow-up with the IRS (Table \ref{tab_summary}).

In addition, we matched the most recent catalog of SDSS type 2 quasars \citep{reye08} against the \spi\ archive and found the following two sources with archival IRS observations: 
\begin{enumerate}
\item SDSS J091345.49+405628.2 ($=$IRAS B09104+4109) -- program 1018 (In-orbit Checkout / Science Verification), PI Armus, z=0.442;
\item SDSS J005009.81-003900.6 ($=$IRAS B00476$-$0054) -- program 105 (Guaranteed Time Observations), PI Houck, z=0.729.
\end{enumerate}
The first of these sources is a well-known luminous type 2 quasar \citep{klei88}, whereas the second source was originally classified as an ultra-luminous infrared galaxy (ULIRG). Both objects show high-ionization optical line ratios, which is why they were selected by the optical procedure described in \citet{reye08}, so we include them in our analysis. 

The ten sources in our targeted program were observed with the first order of the short wavelength low resolution module (SL1; 484 sec exposure for each object) and with the second order of the long wavelength low resolution module (LL2; 488 sec). These observations cover the wavelength range $7.4\micron<\lambda_{obs}<21.3\micron$ with resolution $R \sim  57-127$. We started our analysis from the Basic Calibrated Data (BCD) provided by the \spi\ Science Center (pipeline ver. S13.2.0), which applies dark corrections, flat fields, and droop and nonlinearity corrections, and provides flux calibration. We removed the sky background and most of the hot pixels by a pairwise subtraction of frames at different nod positions. We then extracted positive and negative spectra from the subtracted images with the SMART package \citep{higd04}, using the `interactive/column/No SkySub' method. Additionally, we applied hot pixel corrections to the subtracted two dimensional spectra using IRSCLEAN\footnote{http://ssc.spitzer.caltech.edu/archanaly/contributed/irsclean}. The resulting spectra (4 for SL1 and 8 for LL2 for each object) were averaged into the final spectra shown in Figure \ref{pic1}. When necessary, the spectral errors were estimated from pixel-to-pixel variations by subtracting a smooth model from the spectrum. We found that our reduced spectra are in agreement with IRAC-8.0 and MIPS-24 photometry and conclude that the absolute flux calibration is accurate to better than 15\%.  

We reduced the two archival sources in the same way (pipeline ver. S14.0.0), and they are also shown in Figure \ref{pic1}. The archival observations had total exposures 2-4 times shorter than the targeted observations, since the archival targets are IRAS-detected and therefore are relatively bright in the IR. We have included SL1, LL2 and LL1 observations (total wavelength coverage $7.4\micron<\lambda_{obs}<38.0\micron$). In the source SDSS~J0913+4056 ($=$IRAS B09104+4109), one of the two available LL2 datasets produced fluxes which were systematically lower (by about 5\%) than those from the other LL2 dataset and those extrapolated from the LL1 dataset. We multiplied the fluxes of the discrepant LL2 dataset by a linear function of wavelength to correct for this difference. 

Some of the analysis in this paper makes use of IR photometric data. For the ten targeted sources, the IR photometry is from our Cycle 1 \spi\ program (Zakamska et al., ,in prep.). For SDSS~J0913+4056 ($=$IRAS B09104+4109), we use IRAS catalog fluxes\footnote{taken from the NASA/IPAC Infrared Science Archive, http://irsa.ipac.caltech.edu} ($F_{\nu}[12\micron]=130$ mJy, $F_{\nu}[25\micron]=333$ mJy, $F_{\nu}[60\micron]=525$ mJy) and \spi\ fluxes derived from IRAC observations in program 105. Using aperture photometry, we calculate $F_{\nu}[3.6\micron]=5.0\pm 0.7$ mJy. For SDSS~J0050-0039 ($=$IRAS B00476$-$0054), we use IRAS data from \citet{stan00} ($F_{\nu}[60\micron]=260$ mJy). In all cases where the data overlap, the IR photometry is in agreement with the fluxes derived from the IRS spectra within the uncertainties of the photometric data. 

\section{Analysis of spectral features}
\label{sec_spectra}

\subsection{Identification and detection of spectral features}
\label{sec_features}

\begin{figure}
\epsscale{1.0}
\plotone{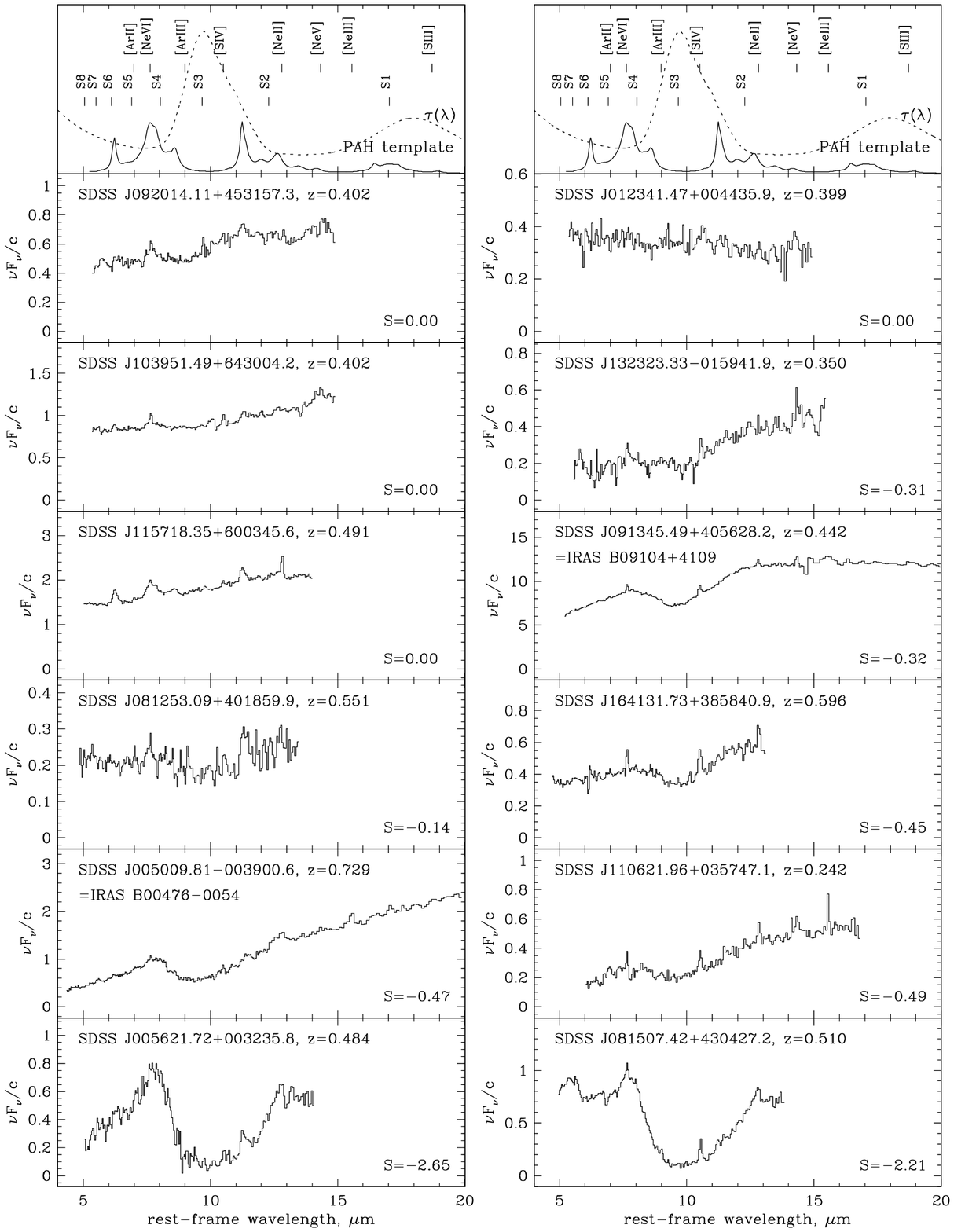}
\figcaption{IRS spectra of type 2 quasars, plotted as $\nu F_{\nu}/c$ in units of mJy/\micron\ vs. rest-frame wavelength. The top-most panels show a PAH template derived from a composite spectrum of galaxies from \citet{smit07}, the Si opacity curve from \citet{kemp04} (dotted curve), and the positions of all the fine-structure atomic emission lines and rotational lines of molecular hydrogen (labeled S1, S2, etc.) that we have considered. Each spectrum includes object ID, redshift and the apparent optical depth of the 9.7\micron\ Si feature derived as described in the text. Objects with detected PAH features are in the left column and those with no PAH features are in the right column. Strength of Si absorption increases from top to bottom. \label{pic1}}
\end{figure}

From Figure \ref{pic1} it is apparent that the IR spectra of type 2 quasars display a wide range of properties. Some are nearly featureless, whereas others are busy with absorption and emission features, some of which are identified in the top panels of Figure \ref{pic1}. The difficulties of placing the continuum to identify detected features, as well as of disentangling blends of emission and absorption are well known for IR spectra (e.g., \citealt{spoo02, spoo04}), and some of our conclusions may suffer from the uncertainties in this rather subjective step. Nevertheless we attempt to minimize these uncertainties or at least obtain their measure by using several different approaches to spectral analysis. 

In the first step, we use PAHFIT \citep{smit07}, a public IDL-based routine that decomposes low-resolution IRS spectra into a combination of starlight (represented by a black body spectrum), featureless continuum (represented by a sum of thermal components at several fixed temperatures), emission features (PAH dust emission features, fine-structure lines of ions, rotational lines of H$_2$), and dust extinction (modeled as a screen of cold silicate dust affecting all other components). The code was developed primarily for the analysis of polycyclic aromatic hydrocarbon (PAH) features in high signal-to-noise ratio spectra of star forming galaxies, so the relative strengths of PAHs are allowed to vary. The advantage of using this code for our objects is that it produces excellent global fits in every case. The disadvantage is that the large number of fit parameters (30-40) makes it hard to assess which of the weak features included in the best PAHFIT model are actually detected. Even for the stronger features, some fit parameters are degenerate, as was illustrated by \citet{spoo02} in the case of the interplay between PAH emission and silicate (Si) absorption. We use PAHFIT as a guide to indicate significant PAH feature detections. 

In the second step, we use a much simpler global fit to the data consisting of a featureless continuum (represented by a second order polynomial) and a PAH template with fixed ratios of feature strengths. The sum of these components is then multiplied by $\exp[-\tau(\lambda)]$, where $\tau(\lambda)$ is the silicate opacity curve, whose normalization is a free parameter which is allowed to be positive or negative. The PAH template that we use is derived using PAHFIT from a composite spectrum of galaxies from Smith et al. (2007; solid line in Figure \ref{pic1}, top), and the Si opacity curve is based on ISO observations of the Galactic center (\citealt{kemp04}; dotted line in Figure \ref{pic1}, top). In some cases we masked the strongest emission lines in the fitting (e.g., [NeVI]7.63\micron, H$_2$ S(3)9.67\micron, [SIV]10.51\micron\ and [NeII]12.8\micron\ in SDSS~J1641+3858). This model has only five free parameters, but still produces good global fits for our spectra. 

Using these two models, we calculate the Si feature strength, $S_{9.7}\equiv\ln(f_{\nu}[9.7\micron]/f_{\nu}[{\rm cont}])$, where $f_{\nu}[9.7\micron]$ is the model flux density at rest-frame 9.7\micron\ and $f_{\nu}[{\rm cont}]$ is the flux density extrapolated from the model continuum outside the feature. In Table \ref{tab_summary} and Figure \ref{pic1}, we quote Si strengths from the second method. The 1$\sigma$ uncertainties shown in Figure \ref{pic_tau} are taken to be half the difference in Si strengths measured by the two methods. 

We tried a range of model Milky Way and Small Magellanic Cloud (SMC) opacity curves from \citet{wein01}, as well as the Sgr A* opacity curve from \citet{kemp04} and the Galactic dust opacity curve from \citet{chia06}. The variation in $S_{9.7}$ measured using different opacity curves amounts to no more than 10\%, although the conversion between IR optical depth and optical extinction or hydrogen column density varies by large factors ($\tau[9.7\micron]=1$ corresponds to $A_V=12.5$ mag for the Milky Way opacity curve and to $A_V=4.7$ mag for the SMC curve). We find that the shape of the 10\micron\ Si feature in type 2 quasars is well-represented by the \citet{kemp04} opacity curve, which we use hereafter. 

In the third step we examine detections of emission lines, such as those from fine-structure transitions in ions (Section \ref{sec_atomic}) and those from rotational transitions of molecular hydrogen (Section \ref{sec_hydro}). As shown in the top panel of Figure \ref{pic1}, we consider all transitions listed in \citet{smit07}, as well as [NeVI]7.63\micron\ and [NeV]14.33\micron. All these lines are expected to be unresolved, so their widths are determined by the instrumental resolution (FWHM$_R=$0.1\micron\ for SL1, 0.14\micron\ for LL2 and 0.34\micron\ for LL1 in the observed frame). For each line, we calculate the error per pixel by fitting a low order polynomial to a cut-out from the spectrum excluding a region 3$\times$FWHM$_R/2.35$ wide on both sides of the line centroid. We then include the region containing the line into the fit and examine the reduced $\chi^2$ with or without a fixed-width Gaussian representing the line to determine the confidence of the detection. Similarly, we examine detections of PAH emission features by fitting a second order polynomial + fixed-width Drude (damped harmonic oscillator) profiles \citep{smit07}. The line fluxes that we obtain from these simple fitting procedures are then compared to PAHFIT line fluxes, allowing us to estimate measurement confidence and uncertainty. For unresolved lines, the uncertainties are in good agreement with those calculated from the measured error per pixel. 

\subsection{Silicates}
\label{sec_si}

Silicates are a major constituent of interstellar and circumstellar dust. Their presence is indicated in many classes of Galactic and extragalactic objects by two strong features, one due to a Si-O stretching mode and peaking at around 10 \micron\ and another due to a O-Si-O bending mode and peaking at around 18\micron\ \citep{knac73}. 

In a recent study of about 200 ULIRGs, Seyfert 2 galaxies, Seyfert 1 galaxies and type 1 quasars, \citet{hao07} find that there exists a strong relationship between the optical classification of these objects and their IR properties. We show their sample together with our type 2 quasars and a compilation of ULIRGs and AGNs from different datasets in a Si strength vs. mid-IR color diagram (Figure \ref{pic_tau}; model lines shown in this Figure are described in detail in Section \ref{sec_models}). For example, the Si feature always appears strongly in absorption in ULIRGs, implying a deeply embedded source of energy with the temperature steeply declining toward the observer \citep{leve07}. The total amount of obscuring material is such that is optically thick even in mid-IR. On the contrary, it was recently confirmed that type 1 quasars show the Si feature in emission \citep{sieb05, hao05b}. The emission feature is thought to be produced by the thin hot layer of dust on the inner side of the obscuring material, where it is illuminated and heated by the central engine. The hot dust emission region is seen without intervening absorption in type 1 quasars. In type 2 quasars from our sample the silicate feature centered at 9.7\micron\ is detected in eight spectra in absorption, and in the remaining four it is not detected, with upper limits of $S_{9.7}<0.1$. As shown in Figure \ref{pic_tau}, the same is true for other samples of type 2 quasars. 

A possible exception is SDSS~J0920+4531 (Figure \ref{pic_0920}). This object has a hump in its mid-IR spectrum centered at about 11\micron, which may be due to Si emission. With this assumption, we estimate the color temperature of this emission -- that is, we fit the continuum-subtracted spectrum $F_{\nu}(\lambda)$ with thermal emission models $B_{\nu}(\lambda,T)\tau(\lambda)$, where $B_{\nu}(\lambda,T)$ is the black-body spectrum with temperature $T$. Our best fit ($T=110$K with \citealt{kemp04} opacity curve) is shown in Figure \ref{pic_0920}. This temperature is significantly lower than those typically derived for type 1 quasars showing Si emission ($\simeq 200$K, \citealt{hao05b}); such a low temperature is required to account for the shift of the feature centroid to long wavelengths. The peak strength of the emission is $S_{11\micron}=0.13$, but there is no flux excess at 9.7\micron\ due to Si emission, so we use $S_{9.7}=0.0$ for this object.

We now investigate the relationship between the strength of Si feature and IR colors. \citet{hao07} find a strong correlation between the depth of the Si feature and the $\log(F_{\nu}[14.5\micron]/F_{\nu}[27.5\micron])$ color in their compilation of ULIRGs, Seyfert galaxies and type 1 quasars, as shown in Figure \ref{pic_tau}. We therefore look for relationships between the values of $S_{9.7}$ and IR colors within our sample of type 2 quasars. To this end, for each object we compile all available IR photometric data from our \spi\ observations (Zakamska et al., in prep.), from IRAS where applicable (Section \ref{sec_data}), and from the IRS spectra themselves (from which we extract fluxes at rest-frame 14.5\micron). We then model the observed spectral energy distribution as a third-order polynomial in the $\log(\lambda)-\log(F_{\nu})$ space, interpolate to calculate model fluxes at rest-frame 3.0, 4.0, 5.5, 8.0, 14.5, 27.5 and 50.0\micron, and compute all 21 rest-frame flux ratios. We then calculate the Spearman rank correlation coefficient ($r_s$) between values of $S_{9.7}$ and these IR colors for our sample of twelve objects. We find that the values of $r_s$ range from 0.2 to 0.75, depending on the color in question, with the most common value around 0.5. The sense of these correlations is that the Si 9.7\micron\ absorption feature is stronger in colder objects, as \citet{hao07} found. Most correlations are rather weak (the values $r_s\simeq 0.5$ imply that there is a $\la 10\%$ chance to find such relationships in uncorrelated datasets), but two are significant at a $>98\%$ level. These are $S_{9.7}$ vs. $F_{\nu}[4.0\micron]/F_{\nu}[8.0\micron]$ and $S_{9.7}$ vs. $F_{\nu}[5.5\micron]/F_{\nu}[50\micron]$; they are shown in the bottom panels in Figure \ref{pic_tau}. In Section \ref{sec_models} we discuss the implications of these findings and the geometrical and physical factors that determine the positions of objects in this diagram.

It would be interesting to determine the relationship between gas column density and dust column density in the obscuring material in AGNs. The column density of neutral gas can be probed directly by X-ray observations. The strength of the Si feature would be a direct measure of the dust optical depth only if the dust were so cold that its emission at 10\micron\ were negligible. In practice dust exists at a range of temperatures and both emits and absorbs in the mid-IR, so that $S_{9.7}$ reflects some combination of optical depth and temperature distribution. \citet{shi06} find a correlation (with a large scatter) between the strength of the Si feature and the neutral gas column density measured using X-ray observations. For 11 out of 12 objects in our sample, pointed {\it Chandra} and {\it XMM} observations are available (Table \ref{tab_summary}). In most cases, only a few photons are detected, and we cannot derive a reliable measurement of the column density from X-ray spectra. Instead, in Figure \ref{pic_lx} we show the strength of the Si feature vs. the X-ray flux of our objects, normalized to their [OIII]5007\AA\ flux presumed to be an isotropic measure of the bolometric luminosity. We calculate X-ray fluxes expected in a type 1 quasar based on the [OIII]5007\AA\ flux. The mean (thick solid line) and the 1$\sigma$ range (dotted lines) of the $L_X/L$[OIII] ratio are taken from \citet{heck05}. We then use the observed X-ray/[OIII]5007\AA\ ratio as a crude measure of the neutral gas column density (the grey lines show the expected ratio for two values of the column density). We see that half of our sample likely has very high column densities, $N_H\ga 10^{24}$ cm$^{-2}$ \citep{ptak06, vign06}, including the two objects with the deepest Si absorption. Although our method of estimating $N_H$ is rather crude, we also see that type 2 quasars occupy a similar range of the $S_{9.7}$-$N_H$ parameter space as do Seyfert 2 galaxies \citep{shi06}.

\subsection{Other absorption features} 

Mid-IR spectra of some deeply absorbed sources, both Galactic (e.g., protostars) and extragalactic (e.g., ULIRGs), display absorption due to organic molecules and inorganic ices (e.g., \citealt{spoo02} and references therein). As we show in Figure \ref{pic_ices}, in SDSS~J0815+4304 we detect a broad absorption feature between 6 and 8\micron\ and within it a pair of narrow features  -- an absorption complex quite similar to the one seen in the ULIRGs IRAS~F00183-7111 \citep{spoo04} and IRAS~08572+3915 \citep{dart07}. 

We identify the two narrow absorption features centered at 6.85\micron\ and 7.27\micron\ with bending (deformation) modes of CH bonds in the CH$_2$ and CH$_3$ subgroups of the aliphatic (chain-like) component of organic molecules within the dust. In Figure \ref{pic_ices}a, we compare the observed absorption with the opacity curve of hydrogenated amorphous carbon measured in the laboratory \citep{dart07}. Using normalizations from \citet{spoo04} and \citet{dart07} and peak optical depths of the two features ($\tau[6.85\micron]=0.19$ and $\tau[6.85\micron]=0.12$), we find column densities of $N_{\rm CH_2}\simeq N_{\rm CH_3}\simeq 6\times 10^{18}$ cm$^{-2}$. This number is quite high; indeed, taking the abundance of carbon to be $3\times 10^{-4}$ relative to hydrogen and assuming that 10\% of it is locked into aliphatic chains \citep{dart07}, we get a neutral hydrogen column density of $N_H\simeq 4\times 10^{23}$ cm$^{-2}$. This value may be compared with $N_H\ga 10^{24}$ cm$^{-2}$ estimated from the lack of an X-ray detection in this object (Figure \ref{pic_lx} and \citealt{vign06}). The difference between the two measurements is indicative of large uncertainties in all the steps of our calculation (such as the conversion between optical depth of the features and $N_{\rm CH_2}$, $N_{\rm CH_3}$, as well as carbon abundance and the fraction of carbon locked in the chains). It is interesting that the peak optical depth at 6.85\micron\ that we derive for SDSS~J0815+4304 (which is a good Compton-thick candidate) is rather modest in comparison with values $\ga1 $ seen in some ULIRGs \citep{spoo02} which are therefore likely to have even higher column densities. 

Water ice absorption and organic molecules (including the stretching modes of CC bonds in ring-like aromatic components) can contribute at wavelengths $\ga 6 \micron$. The precise shape of the opacity curve at $6-8$\micron\ is difficult to determine because it is affected by the choice of the continuum and by assumptions about the strength of the 7.7\micron\ PAH feature. Assuming a power-law continuum and no PAH emission, we derive the opacity curve shown in Figure \ref{pic_ices}b. For the organic component, a broad range of opacity curves has been obtained in the laboratory at these wavelengths \citep{pend02}, but none appears to have the required shape and strength relative to the narrow aliphatic features (a typical opacity curve of laboratory hydrocarbons is shown as the solid line in Figure \ref{pic_ices}b). A water ice feature\footnote{Water ice opacity taken from the database of the Sackler Laboratory for Astrophysics in Leiden, http://www.strw.leidenuniv.nl/$\sim$lab/databases/} with a peak optical depth $\tau[6\micron]=0.30$ can reproduce the absorption at $\lambda\la 6\micron$, but fails at longer wavelengths (dotted line in Figure \ref{pic_ices}b). More importantly, there is no evidence for the stronger 13\micron\ water ice absorption feature in our spectrum ($\tau[13\micron]/\tau[6\micron]\ga 2$ for the water ice opacity curve; \citealt{gera95}), which rules out water ice as the dominant source of opacity at 6\micron. The specific carrier of the excess $6.0-7.5$\micron\ opacity (which has also been observed in ULIRGs, \citealt{spoo02, spoo04, dart07}) remains unidentified. 

\subsection{PAH emission features}
\label{sec_pah}

Many Galactic and extragalactic objects show strong broad emission features at 3--13\micron\ which are thought to originate from small carbonaceous dust particles (\citealt{drai02, drai03}; see top panels of Figure \ref{pic1} for the template of this emission). In contrast to the aliphatic (chain-like) molecules responsible for the absorption bands discussed in the previous section, these emission features are due to polycyclic aromatic hydrocarbon (PAH) molecules in which $\la 50$ carbon atoms are arranged on a hexagonal lattice. The different features arise from different C-C and C-H bending modes \citep{alla89, drai03}, and the precise spectral positions and ratios of the bands depend on which specific molecules are present, on their ionization state and on the spectral shape of the illuminating emission.   

PAH emission is powered by UV photons, and therefore can in principle be powered either by star-forming regions or by the active nucleus. Several arguments favor the hypothesis that PAH emission predominantly traces star formation rather than AGN activity. In particular, 7.7\micron\ PAH emission is spatially extended in nearby AGNs and is suppressed near the nucleus \citep{lefl01}, presumably because of the destruction of PAH molecules by extreme-UV and X-ray photons from the AGN \citep{voit92}. Furthermore, the near-to-far-IR spectral energy distribution of AGNs with PAH emission is accurately represented by a sum of a non-PAH AGN template and a star-burst galaxy template \citep{shi07}. Finally, for quasars with CO emission, the CO-PAH correlation is in agreement with that of star-forming galaxies \citep{shi07}.

Since PAH[7.7\micron] is blended with PAH[8.6\micron], Si absorption, the [NeVI]7.63\micron\ emission line and potentially ice/hydrocarbon absorption as well, we use the strength of the 11.3\micron\ PAH feature as a star formation indicator. Based on the fits described in Section \ref{sec_features}, six objects were judged to have a detected 11.3\micron\ PAH feature. In Table \ref{tab_summary}, we list the average of the fluxes obtained by the two methods (global fitting with PAHFIT and local fitting of a continuum+Drude profile model) and assume that they are accurate to no better than 30\%. We also assume that the PAH emission is not affected by Si absorption (i.e., that it originates from outside the region affected by absorption). If this is not the case, then the 11.3\micron\ luminosities listed in Table \ref{tab_summary} are underestimated. 

\citet{lutz98} proposed a classification scheme in which they used the ratio of peak (continuum subtracted) PAH[7.7\micron] flux density to 7.7\micron\ continuum to distinguish between starburst- and AGN-powered sources, and a similar idea is employed in the `fork diagram' of \citet{spoo07}. A low PAH/mid-IR continuum ratio is characteristic of AGNs, because the spectral energy distributions of star-forming galaxies peak at much longer wavelengths than those of AGNs (Figure \ref{pic_tau}). By the criteria of \citet{lutz98}, all our objects are AGN-dominated in the mid-IR. The PAH/continuum ratio can then serve as a crude estimate of the relative contribution of star formation in the host galaxy to the total IR energy budget. Figure \ref{pic_pah} shows an anti-correlation between the [OIII]5007\AA/[OII]3727\AA\ line ratio and the PAH[11.3\micron]/continuum[27.5\micron] ratio, as well as between [OIII]5007\AA/H$\beta$ and PAH[11.3\micron]/continuum[27.5\micron], which are detected with about 95\% confidence. These anti-correlations suggest that for quasars [OIII]/[OII] and [OIII]/H$\beta$ decrease as the star formation in the host increases, in agreement with classical work on optical line diagnostics \citep{bald81}.

We roughly estimate the total luminosity associated with star formation using the relation $\log(L_{\rm SF}[8-1000\micron]/10^{10}L_{\odot})=1.16\times\log \left(L[{\rm PAH}11.3\micron]/10^{10}L_{\odot}\right)+2.38$ \citep{shi07}. These values are listed in Table \ref{tab_summary}. In Figure \ref{pic_lfsf}, we show the comparison between the distribution of estimated star-formation luminosity in our objects (histograms) and the luminosity function of star formation in other AGNs and in field galaxies from a variety of samples as described in the figure caption. The median star-formation luminosity of our sample ($5\times 10^{11}L_{\odot}$, which includes objects with upper limits on PAH[11.3\micron]) is higher than that of field galaxies and of any other AGN sample. 

As high as the star formation luminosities are, in all but one case they are less than half of the total infrared luminosity obtained by directly integrating the observed IR spectral energy distribution. Thus most of the IR luminosity is due to the dust-reprocessed emission from the quasar (the one exception is discussed below). Models of AGN obscuration (e.g., \citealt{pier92}) show that dust-reprocessed emission is anisotropic even at mid-IR wavelengths, and the luminosities that we obtain in type 2 quasars by integrating the IR emission, multiplying by $4\pi$ and subtracting the star-formation contribution underestimate the quasar bolometric luminosity, by a factor which depends on the viewing angle (see Figure 4 of \citealt{pier92}). When we average model fluxes from \citet{pier92} over all angles such that the central source is not directly seen, we find that this factor ranges from 2.2 to 4.3. This implies that the true bolometric luminosities of the quasars in our sample are likely much larger than the star-formation luminosities of their hosts. Furthermore, in all cases the observed radio luminosity is stronger than what is expected from star formation alone \citep{helo85}, and most of the radio luminosity is due to the quasar, even though all the objects in our sample (except SDSS~J0812+4018) are radio-quiet (based on their position in the $L_{\rm radio}-L$[OIII] diagram; \citealt{zaka04}). 

In the radio-loud type 2 quasar SDSS~J0812+4018, the star-formation luminosity estimated from PAH[11.3\micron] is similar to the total IR luminosity. In this object, the 11.3\micron\ feature appears anomalously strong compared to the 7.7\micron\ one, and therefore the star-formation luminosity is probably overestimated. Although it is usually thought that high-luminosity AGNs destroy PAHs in their vicinity \citep{voit92}, the processes that determine PAH feature ratios are not well-understood, and it is possible that the emission from the quasar itself boosts the 11.3\micron\ feature in this case, as is seen in some low-luminosity AGNs \citep{smit07}. An anomalously strong 11.3\micron\ PAH feature is seen in some other extragalactic objects (e.g., in a buried AGN studied by \citealt{spoo04}).

\subsection{Atomic emission lines}
\label{sec_atomic}

We detect a number of ionized elements through their fine-structure emission lines (the lines that we have considered are enumerated in the top panels of Figure \ref{pic1}). Just as certain optical line ratios are diagnostic of the physical conditions in the medium where they are produced and of the intensity and spectrum of the ionizing radiation field \citep{veil87}, some mid-IR line ratios can also be used as such diagnostics \citep{stur02, grov06}. 

With ionization potentials of 126 eV and 97 eV, the NeVI and NeV species are produced exclusively by the active nucleus, whereas NeII (ionization potential 22 eV) can be excited both by the nucleus and by massive stars. Therefore, the [NeVI]/[NeII] and [NeV]/[NeII] line ratios are diagnostic of the AGN contribution to the line emission. In addition, for a fixed shape of the AGN spectrum, these line ratios increase with increasing ionization parameter (i.e., radiation density, \citealt{grov06}). We detect the [NeVI]7.63\micron\ emission line in 11 out of our 12 spectra. In objects with PAH emission, identification of [NeVI]7.63\micron\ is somewhat complicated by the fact that it is right on top of the PAH[7.7\micron] feature. The median [NeVI]7.63\micron/[NeII]12.8\micron\ ratio in our objects is $\simeq 1.3$, similar to the sample of Seyfert galaxies by \citet{stur02}, where the median is 0.9. The [NeV]14.3\micron\ emission line is also detected in 6 out of the 7 objects in which the wavelength coverage of this line is available. The median [NeV]14.3\micron/[NeII]12.8\micron\ ratio is about 1, similar to that of the sample of \citet{stur02} and somewhat larger than that of Seyfert galaxies from \citet{tomm08} who found a median ratio of 0.5. 

The simple `linear mixing' model of an AGN+starburst spectrum by \citet{stur02} shows that our median value of [NeV]14.3\micron/[NeII]12.8\micron$=1$ roughly corresponds to a 100\% AGN contribution, and our lowest value of this ratio (0.5 in SDSS~J0050-0039) implies an AGN contribution of 50\%. Since [NeV] and [NeVI] are produced in close proximity to the nucleus, they are more affected by extinction in the host galaxy or in the circumnuclear material than [NeII]. Thus, our measured [NeV]/[NeII] and [NeVI]/[NeII] ratios are in fact lower limits, and so is the AGN contribution calculated from the `linear mixing' model. We do not find a strong correlation between the Ne line ratios and the PAH equivalent widths. One likely reason is that a range of ionization parameters is present within our sample, so the dependence of Ne line ratios on the star formation contribution is washed out. 

The one emission line that shows significantly different properties in type 2 quasars and Seyfert galaxies is [SIV]10.51\micron\ (ionization potential 35 eV). This line is strongly detected in 11 of our 12 objects, and the median ratio is [SIV]10.51\micron/[NeII]12.8\micron$\simeq 1.2$ (uncorrected for Si absorption). This ratio is significantly higher than that of the Seyfert galaxies in the sample of \citet{tomm08} (median$\simeq 0.4$) and higher still than that of \citet{stur02} (median$\simeq 0.1$). Perhaps this difference is due to the larger ionization parameter, since the quasars in our sample are significantly more luminous (median $L$(IR)$=10^{45.8}$erg/s) than the Seyfert galaxies in the comparison samples (e.g., $10^{44.3}$ erg/s in \citealt{stur02}).

\subsection{Molecular hydrogen emission lines}
\label{sec_hydro}

Rotational transitions of molecular hydrogen (H$_2$) allow us to directly observe warm ($\ga 200$K) molecular gas and estimate its excitation temperature and mass (e.g., \citealt{rigo02, armu06, higd06, rous07}). The molecule is symmetric, so the transitions are quadrupolar and therefore weak, and their detectability is limited by the low signal-to-noise ratio of our spectra. Furthermore, some of the hydrogen lines blend with other spectral features, which is especially problematic because of our poor spectral resolution. 

Despite these difficulties, we detect several rotational H$_2$ lines in two objects (Table \ref{tab_hydro} and Figure \ref{pic_h2}). The confidence of the detections, uncertainties in fluxes and upper limits are all calculated using simulations of weak unresolved emission lines in noisy spectra. For a line S$(J)$, the transition is from the upper level with angular momentum $J+2$ to the lower level with angular momentum $J$. Under the assumption that emission is purely spontaneous and without correcting for extinction, the line flux $F_J$ can be directly related to the population of the $J+2$ level:
\begin{equation}
N_{J+2}=\frac{4 \pi D_L^2 F_J}{A_{J+2\rightarrow J}(E_{J+2}-E_J)},\label{eq_population}
\end{equation}
where $D_L$ is the luminosity distance to the source, $A$ are the Einstein coefficients (e.g., \citealt{turn77}) and $E_{J+2}-E_J=hc/\lambda$ is the energy of the transition. The rotational energy levels are given by
\begin{equation}
E_J=85.35 {\rm K} \cdot k_BJ(J+1)-0.068{\rm K} \cdot k_BJ^2(J+1)^2,
\end{equation}
where $g_J$ is the degeneracy of the levels, $2J+1$ for even $J$ and $3(2J+1)$ for odd $J$ (assuming equilibrium ortho-to-para ratio) and $k_B$ is the Boltzmann constant \citep{hube79}. In this expression, the first term is the quantum-mechanical analog of the energy of a rotator with angular momentum $J$. As $J$ increases, the energy of the ground vibrational level is changed by rotational stretching, which is taken into account by the second term \citep{land91}. If different levels are populated in a Boltzmann equilibrium, the line ratios are indicative of the excitation temperature which is defined as
\begin{equation}
\frac{N_J}{g_J}\propto \exp\left(-\frac{E_J}{k_BT_{\rm exc}}\right).\label{eq_exc}
\end{equation}

In the spectrum of SDSS~J1641+3858, we detect S(3)9.665\micron, S(4)8.026\micron\ and possibly S(5)6.909\micron, allowing us to calculate an excitation temperature for these transitions. The S(1)17.03\micron\ and S(4) lines are detected in SDSS~J0050$-$0039 with a lower confidence. In Figure \ref{pic_h2}, we show the excitation diagrams for both objects, i.e., we plot $\ln(N_J/g_J)$ from equation (\ref{eq_population}) as a function of the energy of the $J$-th level. In SDSS~J1641+3858, we find an excitation temperature of about 700K, somewhat higher than the temperatures found by \citet{higd06} in a large sample of ULIRGs (300-400K). This difference should not be interpreted as a difference in physical conditions, but rather as a selection effect -- specifically, our lack of longer wavelength coverage precludes us from making observations of lower $J$ transitions which trace colder gas. Indeed, in SDSS~J0050$-$0039 the excitation temperature derived from S(1) and S(4) is lower ($\simeq$420K). 

Using these excitation temperatures we can calculate the mass of the gas. Levels are populated according to equation (\ref{eq_exc}) with the normalization given by equation (\ref{eq_population}). Summing up molecules with all $J$ yields a mass of $3.5\times 10^7M_{\odot}$ in SDSS~J1641+3858 (emitting at temperature 700K) and of $1.8\times 10^9M_{\odot}$ in SDSS~J0050$-$0039 (emitting at temperature 420K). For comparison, in several ULIRGs with five or more detected H$_2$ rotational lines \citep{armu06, higd06}, the gas is found at a range of excitation temperatures, with $\sim 10^6M_{\odot}$ of warm gas ($T\ga 1000$K) and $\sim 10^{9}M_{\odot}$ of cool gas ($T\la 300$K). There could be large amounts of even colder gas which would be invisible in rotational transitions of H$_2$. For example, if one of our objects contains $10^{11}M_{\odot}$ of cold gas with $T\la 150$K, the population of $J\ne 0$ levels is so low that even the lowest energy transition S(0)28.22\micron\ ($J=2\rightarrow J=0$) is well below the typical noise level of our data, as well as of the data by \citet{higd06}. Such cold gas can be traced only through lower energy transitions of other molecules, such as CO(1-0). 

Although H$_2$ rotational line emission is often associated with star formation (and then it is strongly correlated with PAH emission, \citealt{rous07}), there is no evidence for PAH emission in SDSS~J1641+3858. If H$_2$ emission originates in the circumnuclear region (for example, $d\la 100$pc), then the column density of the molecular gas responsible for the observed emission is rather high, $N[{\rm H}_2]\ga 5\times 10^{22}(\frac{d}{100{\rm pc}})^{-2}$ cm$^{-2}$, and yet it maintains an excitation temperature of 700K. As in SDSS~J1641+3858, H$_2$ emission in lower-luminosity AGNs is often in excess of what is expected from star formation in the host galaxy \citep{rous07}. There is at the moment no conclusive model for H$_2$ emission in AGNs. 

\section{Comparison with IR spectra of other type 2 quasars}
\label{sec_compare}

In this section we compare the IR properties of optically selected type 2 quasars with those selected by other methods. We discuss IR-selected (Section \ref{sec_comp_ir}), radio-loud (Section \ref{sec_comp_rad}) and X-ray selected type 2 quasars (Section \ref{sec_comp_x}).

\subsection{IR-selected type 2 quasars}
\label{sec_comp_ir}

Low-redshift ($z\la 1$) quasars show roughly power-law spectral energy distributions in the IRAC bands, unlike nearby star-forming galaxies whose strong PAH emission dominates the 8\micron\ band. The consequent difference in the mid-IR colors allowed \citet{lacy04} and \citet{ster05} to select tens of candidate type 2 quasars, which were then confirmed by optical spectroscopy. \citet{lacy07} present IRS data for some of the objects from that sample; their Si absorption and IR colors are shown in Figure \ref{pic_tau} as cyan squares. They also calculate the total PAH luminosity for the objects in their sample. On the basis of the PAH template described in Section \ref{sec_features}, we estimate that the luminosity of the 11.3\micron\ feature is 0.12 times that of the total PAH luminosity, and therefore the median 11.3\micron\ luminosity of the IR-selected objects is $10^{9.3}L_{\odot}$ -- the same as for our sample.

Combining IRS and HST imaging data for the IR-selected type 2 quasars, \citet{lacy07} find that the depth of the Si feature is correlated with the disk orientation, in the sense that objects that appear more edge-on have a significantly deeper feature. These authors conclude that some of the obscuration occurs in the plane of the host galaxy, rather than in the circumnuclear material on much smaller spatial scales (unless the circumnuclear material is aligned with the plane of the host galaxy). In Figure \ref{pic_orientation} we show both samples (Lacy's and ours) on the $S_{9.7}$ vs. axis ratio diagram. Morphologies and axis ratios are available from HST images for six of our objects \citep{zaka06, armu99}. The IR-selected objects (open squares) tend to show that the Si absorption is stronger for the more inclined galaxies, but the trend disappears when our optically-selected type 2 quasars are included (solid circles). There is a significant difference between the host properties of the IR-selected type 2 quasars and the objects in our sample. The former lie in disk-dominated, disturbed galaxies with dust lanes and active star formation. On the contrary, all our objects are hosted by bulge-dominated galaxies (2/3 of the hosts are ellipticals). Since the IR-selected sample is hosted by disk galaxies, the axis ratio is a good measure of inclination in the IR-selected sample, but not in ours, and the lack of correlation in Figure \ref{pic_orientation} is unsurprising. Additionally, [NeII]12.8\micron/[OII]3727\AA\ ratio is on average 0.8 dex higher in the sample of \citet{lacy07} than in our objects. This difference is suggestive of a higher optical extinction in the host galaxies of the IR-selected objects than in ours, as supported by observations of their optical morphology. 

Why is there such a difference between the host properties of the two samples? The redshift ranges of both samples are similar, and so are the values of Si absorption, IR colors and star-formation luminosities. Optical properties (such as line ratios) are also indistinguishable. The only significant difference seems to be in the luminosities of the two samples. Indeed, the median [OIII]5007\AA\ luminosity of the objects from \citet{lacy07} is $\log (L$[OIII]/$L_{\odot})=8.75$ (calculated from their published [OII]3727\AA\ luminosities using the [OII] vs. [OIII] correlation by \citealt{zaka03}), while it is $\log (L$[OIII]/$L_{\odot})=9.41$ for the objects in our sample. Similarly, the median 27.5\micron\ monochromatic IR luminosity for Lacy's sample is $10^{44.8}$ erg/s, while it is $10^{45.3}$ erg/s for our sample. It appears that there is a qualitative change in the host properties of type 2 quasars occurring at a luminosity bracketed by the two samples. As it is possible that some subtle selection effect leads to the observed difference in host properties, this luminosity dependence clearly warrants further study. 

At higher redshifts ($z \ga 1$), type 2 quasars may be selected via a combination of IR and radio properties. Strong radio emission ensures that the object is an active nucleus (in the `radio intermediate' regime) rather than a star-forming galaxy, while a large mid-IR/near-IR ratio is used to pre-select obscured rather than unobscured objects \citep{mart06}. The IRS spectra of type 2 quasars selected using this method are presented by \citet{mart08}; about half of these objects were confirmed by optical spectroscopy. None of the objects show the 10\micron\ Si feature in emission, and roughly a third appear featureless, similar to our sample. Because of the higher redshift, the IRS spectra of these sources do not cover wavelengths longward of the 10\micron\ Si absorption feature, so we do not show these objects on the Si strength - IR color diagram, but the range of Si absorption in these objects seems to be comparable to that of our sample. The `high-PAH' subsample of the objects from \citet{mart08} has PAH[11.3\micron] luminosities of $10^{10-10.4}L_{\odot}$ -- similar to the highest star-formation luminosities in our sample (Table \ref{tab_summary}). This subsample appears to be well-matched in IR luminosity to the objects in our sample, while the `no-PAH' subsample is 3-4 times more luminous.

\citet{poll08} present a \spi\ spectroscopic study of very luminous ($\nu L_{\nu}[6\micron]>10^{45.6}$ erg/s) high-redshift ($z\ga 2$) obscured quasars. These objects were selected from large-area \spi\ surveys based on their rest-frame NIR colors and a large IR-to-optical ratio, and therefore optical spectroscopy is not available for most sources. Those of the sources that showed 10\micron\ Si absorption made up their sample of 21 obscured quasars. On the basis of this sample, the authors find that obscured quasars with Si absorption constitute more than half of the quasar population even at the highest luminosity end. Featureless IR spectra are common in the optically-selected and IR-selected samples of type 2 quasars (and in the following sections we show that they are common in radio-selected and X-ray selected samples as well), and therefore it is possible that yet more obscured quasars remain to be discovered at these high luminosities and redshifts. 

\subsection{Radio galaxies}
\label{sec_comp_rad}

In Figure \ref{pic_tau} we show Si strengths and IR colors of radio-loud AGNs from the sample of \citet{haas05}, with filled magenta triangles for broad-line radio galaxies and filled blue triangles for narrow-line radio galaxies. Si strengths are estimated from the published spectra by computing $\ln(f_{\nu}[9.7\micron,\mbox{observed}]/f_{\nu}[9.7\micron,\mbox{extrapolated continuum}])$. Again, although a quarter of these objects show spectra with no Si absorption or emission, in the remaining cases the broad-line objects show Si in emission and the narrow-line objects show Si in absorption. The median [OIII]5007\AA\ luminosity of the narrow-line radio galaxies in this sample is $\log L$[OIII]/$L_{\odot}=8.56$, so these objects are roughly matched in luminosity to the IR-selected sample of \citet{lacy07} and are less luminous than ours.

A larger sample of radio-loud AGNs with IRS observations was compiled by \citet{clea07}. More than half of their sample of 28 sources have no detectable Si feature. In the remaining objects, Si appears in emission in radio quasars and in weak absorption in radio galaxies (which in the optical either show narrow lines or no lines at all). There are two exceptions to this trend: 3C343, a radio quasar, shows Si weakly in absorption, and 3C325, a radio galaxy, shows Si weakly in emission. \citet{grim05} find weak broad lines and a red continuum in the optical spectrum of 3C325, and they therefore classify this object as a reddened quasar rather than a radio galaxy. Since the rest-frame UV opacity of dust is much larger than its IR opacity, a small amount of reddening could produce strong UV extinction, while the IR spectrum would be transmitted unaffected. We further comment on the correspondence between the optical classification and the appearance of the Si feature in the next section.

The range of Si absorption appears to be limited in both samples of radio-loud objects, with more than half showing no feature at all and no objects having $S_{9.7}\la -1$. The distributions of $S_{9.7}$ in radio-quiet type 2 quasars (taking our sample and that of \citealt{lacy07}) and in radio galaxies in the sample of \citet{haas05} are different at $>90\%$ level (determined using the Kolmogorov-Smirnov test). One possibility is that the contribution of jet to the mid-IR emission dilutes the spectrum, making the Si feature weaker relative to the total continuum. A larger sample of radio galaxies with well-measured Si strength is needed to investigate this difference in more detail.  

Few, if any, objects in the samples of \citet{haas05} and \citet{clea07} show PAH emission. In a large sample of 3CR objects (including those from \citealt{haas05} and \citealt{clea07}), \citet{shi07} find a median star-formation luminosity of $3\times 10^{10}L_{\odot}$ among objects with detectable PAHs, and $6\times 10^9 L_{\odot}$ when the upper limits on PAHs are accounted for. This level of star formation is lower than in PG quasars or 2MASS AGNs, and more than an order of magnitude lower than that of optically and IR-selected type 2 quasars. 

\subsection{X-ray selected type 2 quasars}
\label{sec_comp_x}

IR spectra of seven X-ray selected type 2 AGNs were presented by \citet{stur06}. A wide range of IR luminosities is represented, from $\nu L_{\nu}[7\micron]=10^{43.5}$ erg/s to $10^{45.4}$ erg/s. Unlike our sample, the IR spectra of these objects are mostly featureless. PAH features are firmly detected in only one object, and in all sources the star-formation luminosity is constrained to be $<$ a few times $10^{10} L_{\odot}$ -- significantly smaller than that of the objects in our sample. None shows the 10\micron\ Si feature in absorption (in one of the seven cases the spectrum does not cover the wavelength range affected by the Si feature). One object (LH 28B) may have Si in emission. This source shows broad lines with a large Balmer decrement in its optical spectrum \citep{lehm00}, so it may be another case of a reddened AGN, similar to 3C325 described above.

One possible reason for the difference in the appearance of the Si feature is that the two samples probe different ranges of column densities. Of the seven objects in the X-ray selected sample, only one (CDF-S 202) has $N_H>10^{24}$ cm$^{-2}$, and that is the object in which the wavelength coverage of the Si feature is not available. The remaining objects have column densities below $10^{23.5}$ cm$^{-2}$. Although we do not have direct measurements of the column density in our sample, more than half of our objects (and in particular, two objects with the deepest Si absorption) are probably Compton-thick, as we discussed in Section \ref{sec_si} and as we show in Figure \ref{pic_lx} \citep{ptak06, vign06}. None of the four type 2 quasars in our sample with $N_H\la 10^{23.5}$ cm$^{-2}$ shows strong Si absorption (top right corner of Figure \ref{pic_lx}). \citet{shi06} also find that deep Si absorption is more likely to occur in Compton-thick sources. Therefore, a bias against Compton-thick sources in X-ray selected samples may result in a bias against deep Si absorption in their IR spectra. 

\section{Si strength vs. IR color diagram}
\label{sec_models}

\subsection{Quasar sequence}

\begin{figure}
\epsscale{1.0}
\plotone{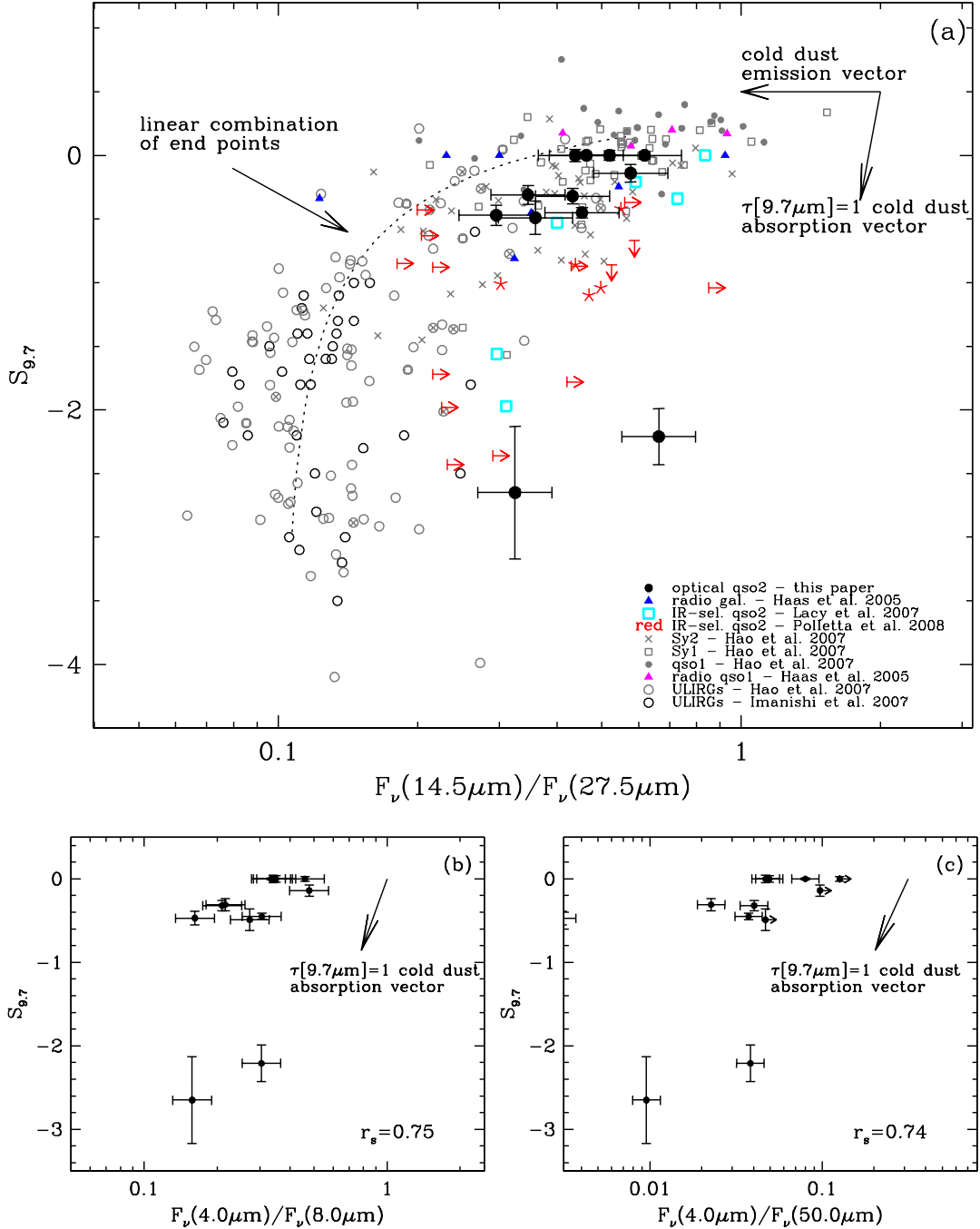}
\figcaption{Empirical relationship between the strength of the 10\micron\ Si feature ($S_{9.7}$) and rest-frame IR colors for different types of objects. Black solid circles with error bars are SDSS type 2 quasars from this paper. There are four objects for which $S_{9.7}=0$ (the upper limits are all $<0.1$). The vertical error bars are from comparison of measures of $S_{9.7}$ by different methods and the horizontal error bars assume a 20\% uncertainty in rest-frame colors (which includes the uncertainty in the photometric measurements and the uncertainty associated with the rest-frame correction). These error bars are representative of the measurements taken from other samples. Panels (b) and (c) show the IR colors which correlate most strongly with $S_{9.7}$ for our sample (both correlations detected at $>98\%$ significance). Additional notation pertains to Section \ref{sec_models}. In panels (a)-(c) we show a `screen of cold dust' absorption vector based on the extinction curve of \citet{kemp04}. In panel (a), we show the cold dust ($T\la 80$K) emission vector which corresponds to doubling the 27.5\micron\ luminosity. The dotted line in panel (a) shows a locus of linear combinations of spectra corresponding to the end points of each curve. For this particular realization, we take the composite spectrum of type 1 quasars from \citet{hao07} as the top right point and the spectrum of IRAS 01166-0844 from \citet{iman07} as the bottom left point. \label{pic_tau}} 
\end{figure}

\begin{figure}
\epsscale{0.5}
\plotone{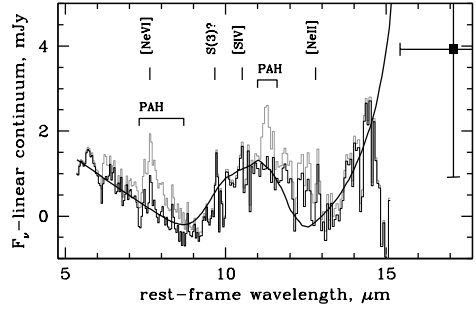}
\figcaption{The spectrum of SDSS~J0920+4531 showing possible Si emission, with PAH features (grey histogram) and without PAH features (black histogram; obtained by subtracting a scaled PAH template from \citealt{smit07}). The black point with error bars is based on our MIPS-24 photometry of this source. A linear continuum (anchored at 8\micron\ and 13\micron) has been subtracted to show deviations more clearly. A similarly normalized spectrum of optically thin dust emission with \citet{kemp04} opacity curve and temperature 110K is shown for comparison. Models with a lower temperature have a narrower Si feature and over-predict long-wavelength continuum; models with a higher temperature have a Si feature shifted to shorter wavelengths. Atomic emission features are indicated (Section \ref{sec_atomic}); the $S(3)$ rotational line of H$_2$ (Section \ref{sec_hydro}) falls right between SL1 and LL2 modules of IRS and is therefore only a tentative detection. \label{pic_0920}} 
\end{figure}

\begin{figure}
\epsscale{0.5}
\plotone{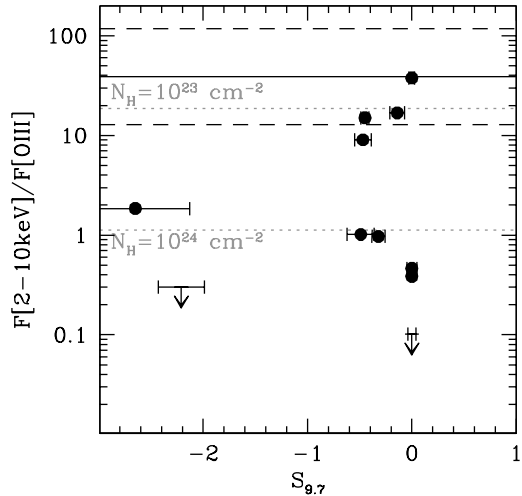}
\figcaption{Strength of Si absorption plotted vs. the ratio of X-ray flux to that of [OIII]5007\AA. The 2-10keV X-ray fluxes have been corrected to the rest-frame. The solid line shows the mean ratio seen in Seyfert 1 galaxies and the dashed lines show the 1$\sigma$ range of this measurement \citep{heck05}. We then take a $\nu F_{\nu}=$const X-ray spectrum and apply photoelectric absorption (S.Davis, private comm.) and Thompson scattering to calculate the X-ray/[OIII]5007\AA\ ratio in presence of line-of-sight gas with hydrogen column density $N_H$ and solar metallicity (grey dotted lines); [OIII]5007\AA\ is assumed to be unaffected by absorption. \label{pic_lx}} 
\end{figure}

\begin{figure}
\epsscale{0.9}
\plotone{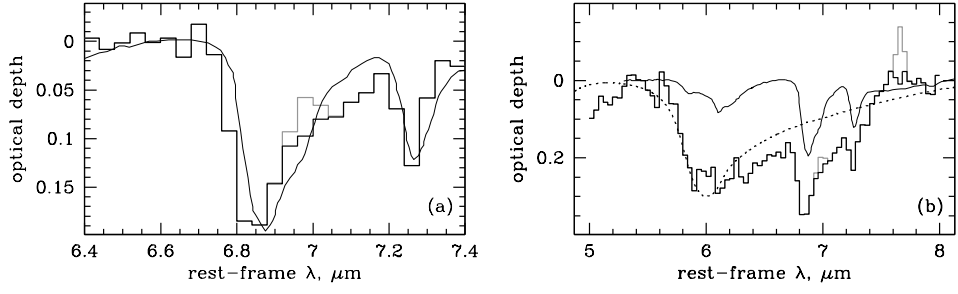}
\figcaption{Optical depth of the absorption complex in SDSS~J0815+4304 (black histogram; grey histogram shows the subtracted [NeVI]7.63\micron\ and [ArII]6.99\micron\ emission lines). Left: a close-up view of the aliphatic features. Right: the opacity of the whole complex when the continuum is assumed to be a power-law between 5.5\micron\ and 7.8\micron. The thin solid line shows a typical laboratory hydrogenated amorphous carbon opacity curve (copied from \citealt{dart07}). The dotted line shows opacity of water ice at $T\la 100$K \citep{gera95}. \label{pic_ices}}
\end{figure}

\begin{figure}
\epsscale{0.5}
\plotone{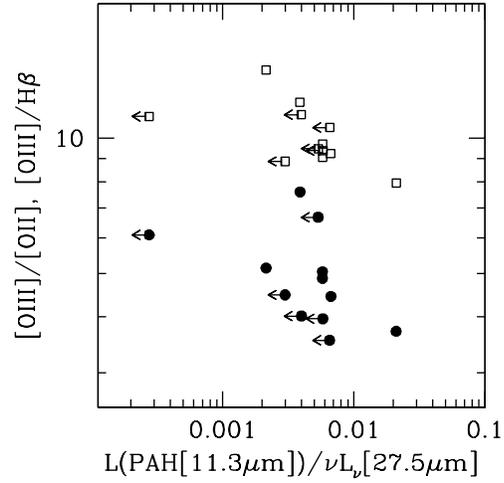}
\figcaption{Optical emission line ratios (open squares: [OIII]5007\AA/H$\beta$; filled circles: [OIII]5007\AA/[OII]3727\AA) vs. the relative contribution of PAH[11.3\micron] to the IR emission. The latter is quantified by the ratio of PAH[11.3\micron] luminosity to the monochromatic luminosity $\nu L_{\nu}$[27.5\micron]. Both optical line ratios tend to decline as the relative PAH contribution increases -- that is, as the relative contribution of star formation increases (anti-correlation is detected at the 95\% confidence level using the rank correlation coefficient; correlation detected both when upper limits are included and when they are excluded). No definitive trends are seen for any other optical or mid-IR line ratios. \label{pic_pah} }
\end{figure}

\begin{figure}
\epsscale{0.5}
\plotone{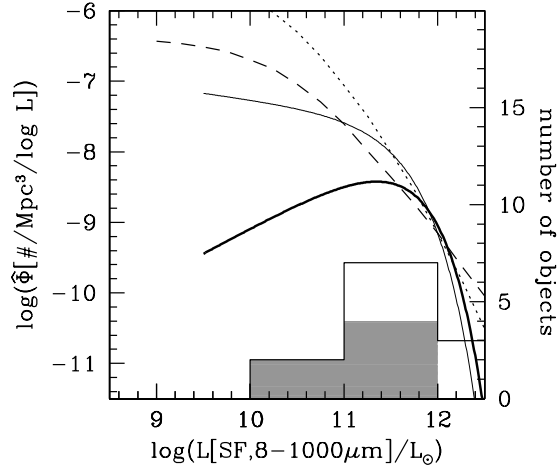}
\figcaption{Distribution of star-formation luminosities in type 2 quasars (right vertical axis; solid line histogram -- all objects; grey shaded histogram -- upper limits) plotted with the same horizontal scale as the luminosity functions (left vertical axis) of star formation in local field galaxies (dotted line, \citealt{lefl05}, arbitrarily normalized), in local Seyfert 1 galaxies (dashed line, \citealt{maio95}, arbitrarily normalized), in $M_B<-21.0$ mag PG quasars (thin solid line, \citealt{shi07}; median star-formation luminosity $6\times 10^{10}L_{\odot}$) and in $M_B<-23.5$ mag PG quasars (thick solid line, \citealt{shi07}; median star-formation luminosity $1.4\times 10^{11}L_{\odot}$). The median star-formation luminosity in type 2 quasars ($5\times 10^{11}L_{\odot}$) is higher than the median star-formation luminosity in any other group of AGNs studied by \citet{shi07}. \label{pic_lfsf} }
\end{figure}

\begin{figure}
\epsscale{1.0}
\plotone{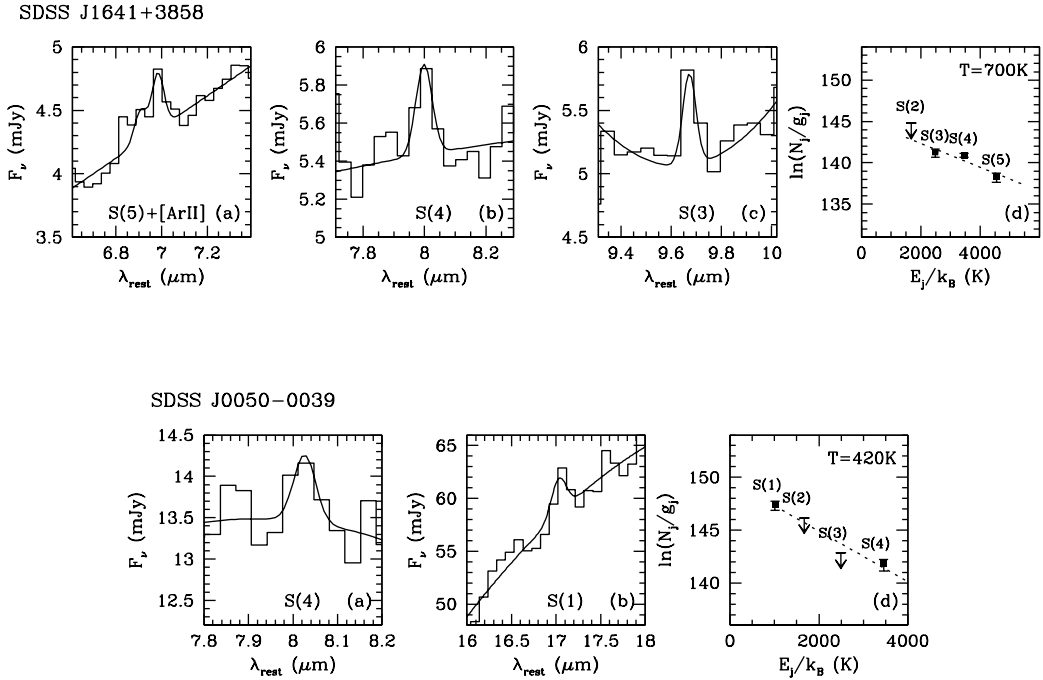}
\figcaption{Top row: rotational lines of molecular hydrogen in SDSS~J1641+3858. (a-c): Solid lines show our best continuum + one Gaussian fits, or two Gaussians in the case of the S(5)+[ArII] blend in panel (a). (d): Excitation diagram with the vertical axis reflecting the actual number of emitting molecules. The excitation temperature is by definition the inverse of the slope of the excitation diagram. Bottom row: tentative H$_2$ line detections and excitation diagram for SDSS~J0050$-$0039. \label{pic_h2}} 
\end{figure}

\begin{figure}
\epsscale{0.5}
\plotone{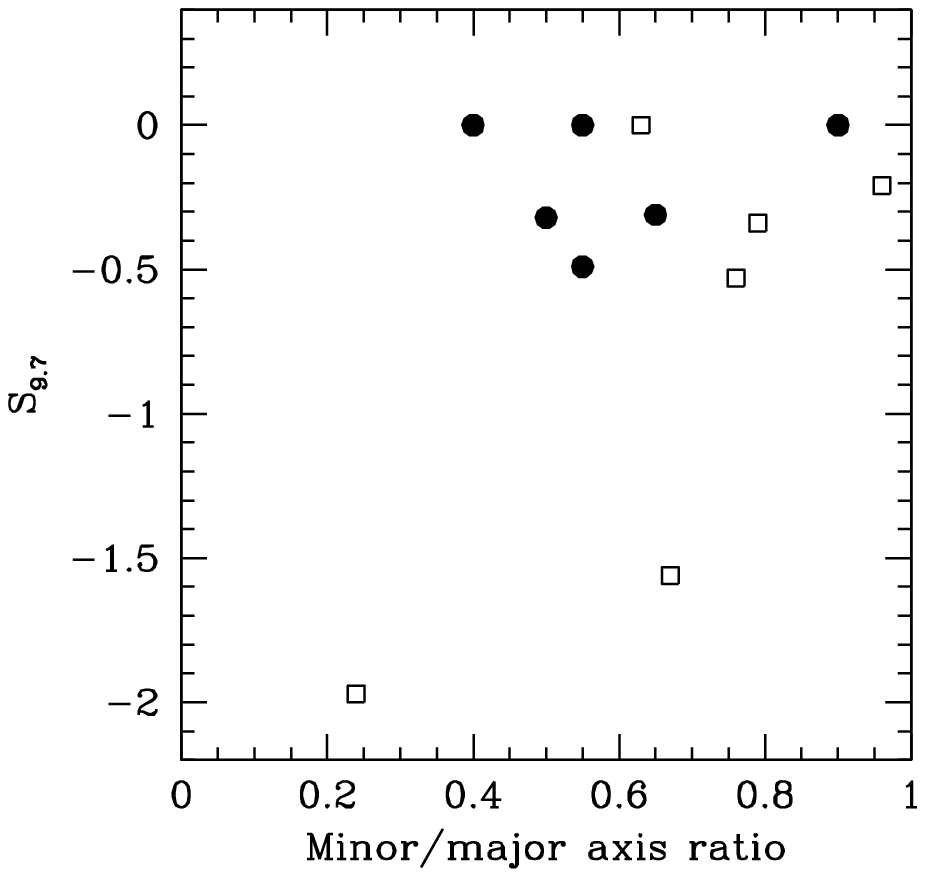}
\figcaption{Si strength vs. axis ratio of the host galaxy. Solid circles are type 2 quasars from our sample. Axis ratios are taken from HST images by \citet{zaka06} for five objects; for SDSS J0913+4056 the ratio is estimated to be 0.5 from images by \citet{armu99}. Open squares are IR-selected type 2 quasars from the sample by \citet{lacy07} who see a trend of increasing Si absorption with increasing ellipticity in their sample. When the objects from our sample are added, the trend disappears. The objects in the sample of \citet{lacy07} reside in dusty disks, whereas the objects in our sample are about 0.5 dex more luminous and reside in bulge-dominated galaxies. \label{pic_orientation}}
\end{figure}

In this section we discuss the position of various types of sources (AGNs and ULIRGs) on the $S_{9.7}$-IR color diagram (Figure \ref{pic_tau}). We start with the type 1/type 2 quasar sequence. In our HST data, some type 2 quasars clearly show conically shaped scattering regions in blue continuum light, with opening angles (projected on the sky) ranging from 5\degree\ to 60\degree\ \citep{zaka06}. If we could determine the inclination of the scattering cones relative to the line of sight, we would know our viewing angle relative to the obscuring material. For example, \citet{zaka06} argued that in SDSS~J1039+6430 the scattering region to the south of the nucleus is oriented toward the observer because the opposing northern scattering region is fainter. If so, we may be able to see down to hotter areas where dust is heated by the quasar, even if not deep enough to see the broad-line region itself. This would be consistent with the absence of a Si absorption feature in this object ($|S_{9.7}|<0.05$). On the other hand, the two objects that have rather symmetric scattering regions in the HST images, SDSS~J1323$-$0159 and SDSS~J1106+0357, may be seen almost exactly edge-on. Both these objects show the Si feature in absorption ($S_{9.7}=-0.16$ and $-0.23$, correspondingly). SDSS~J0920+4531, the only object in our sample with a possible Si emission, is also the only object in which an extended scattering region could not be identified from the HST images \citep{zaka06}. It is likely that the compact blue source in the center of this object is the one responsible for the observed polarization, and it is possible that the scattering cone appears compact because it is viewed close to its axis (almost face-on). Thus, we may be seeing a dependence of Si strength on the viewing angle within the type 2 quasar population. We also saw in the previous section that the appearance of the 10\micron\ Si feature (in emission or in absorption) is well-correlated with the optical classification (type 1 or type 2). These findings imply that the type 1/type 2 quasar sequence in Figure \ref{pic_tau} is driven at least in part by the viewing angle.

This strong relation between the appearance of the Si feature and the viewing angle is rather surprising. First, the broad-line region is much more compact [$\sim 0.05{\rm pc}\cdot\frac{M_{\rm BH}}{10^8M_{\odot}}\left(\frac{{\rm FWHM}_{\rm BLR}}{7000{\rm km\, s}^{-1}}\right)^{-2}$ from the virial argument for bulk motions; cf. reverberation mapping sizes by \citealt{pete93}] than the hot dust which is on scales larger than the region of dust sublimation [$\sim 0.4{\rm pc}\cdot(\frac{L}{10^{46}{\rm erg\, s}^{-1}})^{1/2}\left(\frac{T_d}{1200{\rm K}}\right)^{-2}$ from the heating vs. cooling energy balance for dust particles]. Therefore, there might be plenty of lines of sight for which the broad line region is obscured while the hot dust region is not, with the Si feature appearing in emission, as seen in some geometries in radiative transfer models by \citet{pier92}. However, very few, if any, of the optically-classified type 2 quasars show the feature in emission. The second complication is that the appearance of the Si feature is very sensitive to the geometry of the torus, its total optical depth and the distribution of dust within it. For example, models of Compton-thick tori always show the feature in absorption \citep{pier92}, independent of orientation, because the torus is too optically thick to be heated through in the direction parallel to the axis of symmetry, and large temperature gradients are present along all lines of sight. 

In clumpy tori \citep{nenk02} the Si appearance depends mostly on the distribution and sizes of clumps, rather than on the orientation of the \begin{sloppy}observer\footnote{The library of spectral energy distributions is provided by the CLUMPY group at https://newton.pa.uky.edu/$\sim$clumpyweb/.}\end{sloppy}. When clumpiness is reduced and the volume-filling factor of clumps is increased, Si shows a stronger dependence on the observer's orientation \citep{scha08}. Therefore, the observed relation between the optical classification and the appearance of the Si feature may prove useful in constraining clumpy models of AGN obscuration. Another characteristic of clumpy models of \citet{nenk02} is that the shape of the Si feature in the final spectrum is often significantly different from the shape of the feature in the opacity curve, due to a superposition of emission from regions with different temperatures. This diversity is not seen in those of our spectra that show Si in absorption, while the very low color temperature of the possible Si emission in SDSS~J0920+4531 is a challenge to all existing models. The high fraction of featureless spectra, the centroid and the shape of the 10\micron\ feature, the relative strengths of the 10\micron\ and the 18\micron\ features (not available for most our objects, but employed in a ULIRG study of \citealt{siro08}) and the relationship between Si strength and the observed $N_H$ \citep{shi06} may prove to be useful diagnostics of AGN models.  

The observed mid-IR $F_{\nu}[14.5]/F_{\nu}[27.5]$ color of all quasars in Figure \ref{pic_tau} has a median of 0.5$\pm$0.2 (with 60\% of objects falling within this range). This is not well reproduced by either smooth or clumpy models, which typically predict $F_{\nu}[14.5]/F_{\nu}[27.5]>1$. However, the models do predict that this color should be rather similar in type 1 and type 2 quasars, as observed (type 1 quasars in Figure \ref{pic_tau} have a ratio of $0.6^{+0.3}_{-0.2}$, while type 2 quasars have a ratio of $0.4\pm0.1$). In order to bring the model color in agreement with observations, one could postulate a cold dust component in the host galaxy. As long as the temperature of this component is $\la 80$K, its flux density at 14.5\micron\ is more than an order of magnitude smaller than that at 27.5\micron\ (and it makes an even smaller contribution to the wavelengths of the Si feature). We can estimate the minimal mass of this component by assuming that it is an optically thin emitter [$F_{\nu}(\lambda)\propto B_{\nu}(T,\lambda)\tau(\lambda)$] and by requiring that it boosts the luminosity at 27.5\micron\ (taken to be $10^{45}$ erg/s) by a factor of two (see Figure \ref{pic_tau}), which is sufficient to bring the $F_{\nu}[14.5]/F_{\nu}[27.5]$ from 1 to 0.5. The Milky Way opacity curve from \citet{drai02} is used to convert between optical depth and hydrogen column density ($N_H\simeq 1.2\times 10^{23}\tau[27.5\micron]$ cm$^{-2}$). The minimal hydrogen mass is $3\times 10^{8}M_{\odot}$ if the cold component has $T=80$K, but the required mass grows rapidly as the assumed temperature is decreased (e.g., $1.1\times 10^{10}M_{\odot}$ at $T=50$K), since more mass is required to produce a given amount of flux at a colder temperature. The mass may be underestimated by another factor of 4 if the gas-to-dust ratio is similar to the SMC value rather than to the Milky Way value \citep{drai02}. 

Because the radiative transfer models of circumnuclear obscuration in AGNs predict similar colors for type 1 and type 2 objects, it is impossible to distinguish between multi-temperature circumnuclear obscuration and intervening cold absorption on the basis of IR spectra alone. Indeed, the opacity curve of dust is rather wavelength-independent (outside the strong Si features at 10 and 18\micron), so intervening cold absorption does not change the colors in a significant way (see the `screen of cold dust' absorption vector in Figure \ref{pic_tau}a). In other words, the IR spectra of type 2 quasars look just like they would if there was a screen of cold dust in front of a type 1 quasar. The correlations between $S_{9.7}$ and IR colors found in our sample of 12 objects shown in Figures \ref{pic_tau}b,c also seem consistent with the `screen of cold dust' model. Our HST observations of scattering cones strongly suggest obscuration close to the nucleus \citep{zaka06}, so perhaps the outer parts of the obscuring torus are acting as this screen, while contributing emission at long wavelengths as described in the previous paragraph. Obscuration by intervening cold dust in the host galaxy seems to be important in the less luminous sample of \citet{lacy07}, as we described above.

In type 1 quasars, there is a well-defined maximum value of the strength of Si emission, around $S_{9.7}=0.5$. To understand the origin of this maximum, we use DUSTY\footnote{The DUSTY code is available at http://www.pa.uky.edu/$\sim$moshe/dusty/.} to calculate emission of optically-thick slabs illuminated by the quasar spectrum on one side. All incident radiation is absorbed in a thin layer near the illuminating source. Some of it is transmitted through the slab, while some of it is re-emitted toward the source. The spectrum of this `reflected' component is almost independent of the optical depth, but depends somewhat on the assumed temperature at the illuminated surface, with Si strength $S_{9.7}=0.9-1.6\times\log(T_{\rm in}/1200{\rm K})$. So if the temperature on the illuminated surface is close to the dust sublimation temperature ($\sim 1200$K), the model reproduces the observed maximum Si strength. The observed emission from quasars includes emission from the much colder outer parts of the obscuring material and thus is colder than the predicted reflected emission from the slab ($F_{\nu}[14.5]/F_{\nu}[27.5]\simeq 2$ at $T_{\rm in}=1200$K). 

\subsection{Seyfert galaxies and ULIRGs}

The Seyfert galaxies in the sample of \citet{hao07} show a range of Si absorption and emission similar to that of the higher-luminosity quasars, as well as similar IR colors ($F_{\nu}[14.5]/F_{\nu}[27.5]=0.44^{+0.14}_{-0.20}$). However, the optical classification of Seyferts does not correlate with the appearance of the Si feature nearly as well as it does for quasars. In particular, more than 50\% of Seyfert 1 galaxies show Si in absorption, while only 10\% of type 1 quasars do. One possible explanation for this difference is that even a modest amount of star formation in a Seyfert galaxy can significantly modify its IR spectrum, since the nucleus is not all that luminous. This possibility is illustrated with a dotted line in Figure \ref{pic_tau}, where we show a locus of linear combinations of quasar and ULIRG spectra to mimic an AGN in a strongly star-forming galaxy. We see that when a star formation contribution is increased, the IR colors rapidly become much colder, contrary to what is observed. Indeed, Seyfert 2 galaxies from the sample of \citet{hao07} have a median color $F_{\nu}[14.5]/F_{\nu}[27.5]=0.38$ and a median Si strength $S_{9.7}=-0.41$, so they are warmer than the composite quasar+ULIRG objects on the dotted curve. Another possible explanation for the difference between IR properties of Seyfert galaxies and quasars is that there is some systematic geometric or physical difference between the two classes because of their different bolometric luminosity. For example, it is possible that the opening angle of obscuration or the degree of clumpiness depend on luminosity, resulting in different relationships between the IR properties and the viewing angle in Seyferts and quasars. Further theoretical work may clarify this interesting difference. 

Type 2 quasars are significantly warmer than are ULIRGs for a given strength of Si absorption. Radiative transfer models show that this difference is not due to the different shapes of the illuminating spectra, since all incident emission is absorbed within a thin layer close to the source \citep{leve07}. The total optical depths are also thought to be comparable in quasars and ULIRGs. Instead, the differing IR colors probably reflect different geometric distributions of dust in quasars and ULIRGs. If the energy sources in ULIRGs (starbursts or AGNs) are completely embedded (that is, covered by obscuring material along all lines of sight), we would not see any warm dust. Using DUSTY, we explore optically thick spherical shells illuminated on the inside for a range of shell geometries, optical depths and mass distribution, following \citet{leve07}. We find that the strength of Si absorption seen in ULIRGs is reproduced by models with IR optical depth, $\tau[9.7\micron]$, of order a few. There is a trend of colder IR colors for less centrally concentrated shells, but all model colors (for reasonable Si strengths) are warm, with $F_{\nu}[14.5]/F_{\nu}[27.5]$ between 0.1 and 1. It appears that a significant fraction of ULIRGs (those with colder colors) require a cold dust component just as AGNs do.

\section{Conclusions}
\label{sec_conc}

We have conducted \spi\ IRS spectroscopy of ten type 2 quasars and obtained spectra of two more objects from the \spi\ archives. The sample was selected from the SDSS spectroscopic database on the basis of optical emission line properties and was required to have a minimal [OIII]5007\AA\ luminosity of $10^9L_{\odot}$ and a minimal IR flux to obtain good quality spectra, but was otherwise unbiased with regard to any multi-wavelength properties. In particular, the sample appears to represent a wide range of column densities along the line of sight, and half of the objects are likely to be Compton-thick. Optically-selected type 2 quasars show a diversity of IR spectral properties. Most objects show high-ionization fine structure emission lines such as [NeVI]7.63\micron, essentially ruling out star formation as the dominant energy source. The 10\micron\ Si feature appears in absorption in eight of the objects with a large range of strengths, and in another three the feature is not detected. In the one remaining object (SDSS~J0920+4531) the feature may have been detected in emission, peaking at 11\micron, which corresponds to an unusually low color temperature of $\simeq 110$K. 

Half of the sample show PAH features. One type 2 quasar (SDSS~J0815+4304) shows absorption due to organic carbonaceous grains and possibly water ice absorption. We detect three rotational lines of molecular hydrogen in one object (SDSS~J1641+3858), with excitation temperature $T_{\rm exc}=700$K and mass $M=3.5\times 10^7M_{\odot}$. Molecular hydrogen is also detected (with a lower confidence) in SDSS~J0050$-$0039.  

The median IR star-formation luminosity of type 2 quasar host galaxies inferred from PAH emission is $5\times 10^{11}L_{\odot}$, but the bolometric output of our objects is dominated by the obscured quasars. These levels of star formation are similar to those found by \citet{lacy07} and \citet{mart08} in other samples of type 2 quasars. The high star-formation luminosities of type 2 quasars -- much higher than in field galaxies -- suggest that quasar activity is associated with intensified star formation in its host galaxy, as suggested by merger-driven models of quasar evolution \citep{hopk06}. Moreover, the median star-formation luminosity that we find is higher than that of type 1 quasar host galaxies, $6\times 10^{10}-1.4\times 10^{11}L_{\sun}$, depending on the quasar luminosity \citep{shi07}. This trend supports the suggestion that the obscured fraction evolves with time, in the sense that the probability to see the quasar as obscured decreases with time after the merger \citep{hopk06}. In this case type 2 quasars would appear on average at an earlier stage of the merger and therefore with a higher star-formation rate than type 1 quasars, as observed. 

Objects in our sample occupy luminous bulge-dominated or elliptical hosts, whereas IR-selected type 2 quasars from the sample of \citet{lacy07} inhabit dusty disk galaxies. The two samples have different luminosities, with the objects from our sample being about 0.5 dex more luminous ($\nu L_{\nu}[27\micron]=10^{45.3}$ erg/s in our sample vs. $10^{44.8}$ erg/s in the sample of \citealt{lacy07}). \citet{hopk08} argue that at $z\ll 1$ most AGN activity at low luminosities is due to gas instabilities in the host galaxy, while high luminosity quasars are fueled in major mergers which result in formation of bulges. One possible explanation for the observed difference between host properties is that the two samples bracket a transition between different quasar fueling mechanisms. 

Both optically-classified type 2 and type 1 quasars often show spectra with no sign of the 10\micron\ Si feature. The fraction of such objects depends on how the sample was selected; overall, about a third of all type 1 and type 2 quasars discussed in this paper do not show the feature. Most quasars that show Si in absorption are optically classified as type 2, while those with Si emission are type 1. Some models of clumpy AGN tori show that the appearance of the feature is primarily determined by the geometrical distribution of the material and its total optical depth, while the viewing angle acts as a secondary parameter. Therefore, the observed correlation between the optical classification and the appearance of the IR spectrum may be a useful discriminator of models of AGN obscuration. We further find that the correlation is significantly weaker in Seyfert galaxies than it is in quasars -- specifically, half of Seyfert 1 galaxies show the 10\micron\ Si feature in absorption. This suggests that either the geometry or the physical properties of obscuring material vary with luminosity.  

\acknowledgments
N.L.Z. is supported by the \spi\ Space Telescope Fellowship provided by NASA through a contract issued by the Jet Propulsion Laboratory, California Institute of Technology. L.G. is grateful to the Institute for Advanced Study and the  Princeton University Observatory for their  hospitality and financial support during the visit. L.G. also acknowledges the support of CONACyT, M\'exico. M.A.S. acknowledges the support of NSF grant AST-0707266. We would like to thank Bruce Draine, Brent Groves, Martin Haas, Lei Hao, Philip Hopkins, Mark Lacy, Yong Shi, Henrik Spoon and the anonymous referee for the useful comments; and Lei Hao, Masatoshi Imanishi and Edward Pier for providing data and models in electronic form. 

This work is based on observations made with the Infrared Spectrograph (IRS) onboard \spi\ Space Telescope, which is operated by the Jet Propulsion Laboratory, California Institute of Technology under a contract with NASA. The IRS is a collaborative venture between Cornell University and Ball Aerospace Corporation funded by NASA through the Jet Propulsion Laboratory and Ames Research Center. The SMART software was developed by the IRS Team at Cornell University and is available through the \spi\ Science Center at Caltech.

Funding for the SDSS and SDSS-II has been provided by the Alfred P. Sloan Foundation, the Participating Institutions, the National Science Foundation, the U.S. Department of Energy, the National Aeronautics and Space Administration, the Japanese Monbukagakusho, the Max Planck Society, and the Higher Education Funding Council for England. The SDSS Web Site is http://www.sdss.org/.

\clearpage
\begin{landscape}
\begin{deluxetable}{ccccccccccc}
%\rotate
\tabletypesize{\footnotesize}
\tablecaption{Data summary}
\tablewidth{0pt}
\tablehead{Object ID & $z$ & $\log\frac{L{\rm [OIII]}}{L_{\odot}}$ & $S_{9.7}$ & $\log L_X$, & $\log L[14.5\micron]$, & $\log L[27.5\micron]$, & $\log L$[1.4GHz], & PAH 11.3\micron, &$L_{\rm SF}$, &$\log L$(IR),\\
 & & & & erg/s (ref.) & erg/s & erg/s & erg/s & $\log L/L_{\odot}$ & $\log L/L_{\odot}$ & erg/s }
\startdata
SDSS J005009.81$-$003900.6  & 0.728& 10.06 & -0.47 & 44.6 (3) & 46.09 & 46.37 & 41.0 & 10.1 & 12.5 & 46.4 \\
SDSS J005621.72+003235.8  & 0.484&  9.25 & -2.65 & 43.1 (1) & 45.25 & 45.37 & 40.9 & 9.5 & 11.9 & 45.6 \\
SDSS J012341.47+004435.9  & 0.399&  9.14 & 0.00 & 44.3 (1) & 44.75 & 44.65 & 40.9 & $<$8.8 & $<$11.0 & 45.0 \\
SDSS J081253.09+401859.9  & 0.551&  9.39 & -0.14 & 44.2 (1) & 44.99 & 44.95 & 43.2\tablenotemark{*} & 9.7 & 12.0\tablenotemark{*} & 45.3 \\
SDSS J081507.42+430427.2  & 0.510&  9.44 & -2.21 & $<$42.5 (3) & 45.37 & 45.31 & 40.8 & $<$9.2 & $<$11.5 & 45.8 \\
SDSS J091345.49+405628.2  & 0.441& 10.33 & -0.32 & 43.9 (1) & 46.48 & 46.48 & 40.8 & $<$9.3 & $<$11.6 & 46.7 \\
SDSS J092014.11+453157.3  & 0.403&  9.15 & 0.00\tablenotemark{**} & 42.4 (1) & 45.10 & 45.13 & $<$39.8 & 9.4 & 11.7 & 45.5 \\
SDSS J103951.49+643004.2  & 0.402&  9.43 & 0.00 & 42.6 (1) & 45.33 & 45.29 & $<$39.8 & 9.3 & 11.6 & 45.7 \\
SDSS J110621.96+035747.1  & 0.242&  9.01 & -0.49 & 42.6 (1) & 44.44 & 44.43 & $<$39.4 & $<$8.4 & $<$10.6 & 44.5 \\
SDSS J115718.35+600345.6  & 0.490&  9.61 & 0.00 & $<$42.2 (3) & 45.78 & 45.78 & 40.2 & 10.0 & 12.3 & 46.3 \\
SDSS J132323.33$-$015941.9  & 0.350&  9.27 & -0.31 & n/a & 44.74 & 44.81 & 39.8 & $<$9.0 & $<$11.3 & 45.1 \\
SDSS J164131.73+385840.9  & 0.596& 10.04 & -0.45 & 44.8 (2,3) & 45.43 & 45.46 & 40.8 & $<$9.6 & $<$12.0 & 45.8 \\
\enddata
\tablecomments{All luminosities are K-corrected to the rest-frame. $L_X\equiv 4 \pi D_L^2 F_X$[2-10keV] (uncorrected for absorption); $L[\lambda]\equiv 4\pi D_L^2 \nu F_{\nu}[\lambda]$. In the last column, $L({\rm IR})=4 \pi D_L^2 F({\rm IR})$, where $F$(IR) is the integral of the IR spectral energy distribution from \spi\ photometry (Zakamska et al., in prep.). PAH luminosities are given uncorrected for absorption; star-formation luminosities are derived from PAH luminosities as described in Section \ref{sec_pah}. Typical uncertainties are 0.1 dex for IR, 0.05 dex for radio and [OIII]5007\AA\ luminosities. Radio data are from FIRST survey \citep{beck95}. Rest-frame 2$-$10 keV X-ray fluxes are obtained by fitting an absorbed power-law spectrum to the data. When only a few counts are detected, a $\Gamma=1.8$ power-law spectrum is assumed. For $L_X\la 10^{43}$, $L_X$ is determined to no better than 0.3 dex because of the Poisson noise and uncertainties in the spectral shape. References for X-ray luminosities: (1) -- unpublished (Ptak et al., in preparation, based on our recent {\it Chandra} and {\it XMM} programs); (2) -- \citet{ptak06}; (3) -- \citet{vign06}. }
\tablenotetext{*}{SDSS~J0812+4018 is the only radio-loud object; the rest are radio-quiet, given their position in the $L$[OIII]/$L_{\rm radio}$ diagram \citep{zaka04}. The star-formation luminosity in this object is likely over-estimated, see Section \ref{pic_pah}.}
\tablenotetext{**}{SDSS~J0920+4531 possibly shows the Si feature in emission, see Figure \ref{pic_0920}. If this is a correct interpretation, the peak strength of this feature is $S_{11\micron}=0.13$; however, there is no excess flux from this emission at 9.7\micron, and therefore we list $S_{9.7}=0$.}
\label{tab_summary}
\end{deluxetable}
\clearpage
\end{landscape}

%\clearpage
\begin{deluxetable}{cccccc}
\tablewidth{0pt}
\setlength{\tabcolsep}{0.03in}
\tablecaption{Molecular hydrogen lines}
\tablehead{ & S(1) flux, & S(2) flux, & S(3) flux, & S(4) flux, & S(5) flux, \\
Object & $\times 10^{-15}$ erg/s & $\times 10^{-15}$ erg/s & $\times 10^{-15}$ erg/s & $\times 10^{-15}$ erg/s & $\times 10^{-15}$ erg/s  \\
name & (17.04\micron) & (12.28\micron) & (9.67\micron) & (8.03\micron) & (6.91\micron) }
\startdata
SDSS J0050$-$0039 & 5$\pm$2 & $<$5 & $<$3 & 1.4$\pm$0.7 & \nodata \\
%SDSS J0920$+$4531 & n/a & $<$2.7 & 2.5$\pm$0.7 & $<$2.8 & \nodata \\
SDSS J1641$+$3858 & n/a & $<$2.1 & 1.0$\pm$0.5 & 0.9$\pm$0.2 & 0.6$\pm$0.3
\enddata
\tablecomments{All upper limits are at 95\% confidence level (in the sense that 95\% of lines with that flux would have been detected at 95\% confidence level given the local signal-to-noise ratio). Spectral coverage of S(1) is not available for SDSS~J1641+3858. We do not give upper limits for S(5) for SDSS J0050$-$0039 because it is blended with [ArII]6.985\micron, and in SDSS~J1641$+$3858 it is only tentatively detected. There is also a possible detection of S(3) in SDSS~J0920+4531 (Figure \ref{pic_0920}); in this case, the line falls in the high noise wavelength regime right between the SL1 and LL2 modules of the IRS, so we do not provide quantitative measurements.}
\label{tab_hydro}
\end{deluxetable}

\clearpage

%\clearpage

%\clearpage

%\clearpage

%\clearpage

%\clearpage

%\clearpage

\end{document}